%% file: nondetection_debris.tex
\documentclass[fleqn,usenatbib]{mnras}

\usepackage{newtxtext}

\usepackage[T1]{fontenc}

\DeclareRobustCommand{\VAN}[3]{#2}
\let\VANthebibliography\thebibliography
\def\thebibliography{\DeclareRobustCommand{\VAN}[3]{##3}\VANthebibliography}

\usepackage{graphicx}	
\usepackage{amsmath}	
\usepackage{easy-todo}

\usepackage{amssymb}

\usepackage{newtxmath}

\newcommand{\kms}{\,km\,s$^{-1}$}
\newcommand{\hcop}{HCO$^+$}
\newcommand{\cch}{C$_{2}$H}

\title[Nature of gas in CO-rich debris discs]{Lack of other molecules in CO-rich debris discs: is it primordial or secondary gas?}

\author[Smirnov-Pinchukov et al.]{
Grigorii V. Smirnov-Pinchukov,$^{1}$\thanks{E-mail: smirnov@mpia.de}
Attila Mo\'or,$^{2,4} \thanks{E-mail: moor.attila@csfk.mta.hu}$
Dmitry A. Semenov,$^{1,3}$
P\'eter \'Abrah\'am,$^{2,4}$\newauthor
Thomas Henning,$^{1}$
\'Agnes K\'osp\'al,$^{2,1,4}$
A. Meredith Hughes,$^5$ and
Emmanuel di Folco$^{^6}$. 
\\
$^{1}$Max Planck Institute for Astronomy, K{\"o}nigstuhl 17, D-69117 Heidelberg, Germany\\
$^{2}$Konkoly Observatory, Research Centre for Astronomy and Earth Sciences, E\"otv\"os Lor\'and Research Network (ELKH), H-1121 Budapest,\\ Konkoly Thege Mikl\'os \'ut 15-17., Hungary \\
$^{3}$Department of Chemistry, Ludwig Maximilian University, Butenandtstr. 5-13, D-81377 Munich, Germany \\
$^{4}$ELTE E\"otv\"os Lor\'and University, Institute of Physics, P\'azm\'any P\'eter s\'et\'any 1/A, H-1117 Budapest, Hungary\\
$^{5}$Astronomy Department and Van Vleck Observatory, Wesleyan University, 96 Foss Hill Drive, Middletown, CT 06459, USA\\
$^{6}$Laboratoire d’Astrophysique de Bordeaux, Univ. Bordeaux, CNRS,
B18N, allée Geoffroy Saint-Hilaire, 33615 Pessac, France\\
}

\date{Accepted 2021 October 26. Received 2021 October 4; in original form 2021 August 3}

\pubyear{2021}

\begin{document}
\label{firstpage}
\pagerange{\pageref{firstpage}--\pageref{lastpage}}
\maketitle

\begin{abstract}
The nature of the gas in CO-rich debris discs remains poorly understood, as it could either be a remnant from the earlier Class~II phase or of secondary origin, driven by the destruction of icy planetesimals.
The aim of this paper was to elucidate the origin of the gas content in the debris discs via various simple molecules that are often detected in the less-evolved Class~II discs.  
We present millimetre molecular line observations of nine circumstellar discs around A-type stars: four CO-rich debris discs (HD 21997, HD 121617, HD 131488, HD 131835) and five old Herbig Ae protoplanetary discs (HD 139614, HD 141569, HD 142666, HD 145718, HD 100453). 
The sources were observed with the  Atacama Large Millimeter/submillimeter Array (ALMA) in Bands~5 and 6 with 
1--2{\arcsec} resolution. The Herbig Ae discs are detected in the CO isotopologues, CN, HCN, \hcop{}, \cch, and CS lines. In contrast, only CO isotopologues are detected in the debris discs, showing a similar amount of CO to that found in the Herbig Ae protoplanetary discs. 
Using chemical and radiative transfer modelling, we show that the abundances of molecules other than CO in debris discs are expected to be very low. We consider multiple sets of initial elemental abundances with various degrees of H$_2$ depletion. 
We find that the \hcop{} lines should be the second brightest after the CO lines, and that their intensities strongly depend on the overall CO/H$_2$ ratio of the gas. However, even in the ISM-like scenario, the simulated \hcop{} emission remains weak as required by our non-detections.
\end{abstract}




\section{Introduction} \label{sec:introduction}

Recent line observations at millimetre wavelengths, mostly carried out
with the ALMA interferometer, have revealed the presence of CO gas around
about one and half dozen main-sequence stars known to harbour optically thin circumstellar debris dust material \citep{hughes2018}.
The observed dust and gas are at least partially co-located at a radial distance of tens of 
astronomical units (au) 
from the host stars. The derived CO masses display a large spread of at least five orders of magnitude, with roughly bimodal distribution.
While about half of the sample has a low CO mass of $\lesssim$10$^{-4}$\,M$_\oplus$, there are at least six discs 
 with the CO mass $>$0.01\,M$_\oplus$, a value that overlaps with that 
of the more-evolved, less massive protoplanetary discs. The CO-rich debris discs
are found exclusively around young (10--50\,Myr) A-type stars 
\citep{kospal2013,moor2017,moor2019}. 
The host stars of discs with lower CO content show a greater diversity: their spectral types range between A and M, and though this subsample is also predominantly young \citep[10--50\,Myr,][]{dent2014,lieman-sifry2016,marino2016,matra2019}, it includes two older systems, Fomalhaut and $\eta$\,Crv with ages of 0.44\,Gyr and 1--2\,Gyr, respectively  \citep{marino2017,matra2017}.

Dust grains in debris discs are thought to be derived from collisions of larger solids up to the size of planetesimals \citep{wyatt2008,hughes2018}. The emerging second-generation small dust particles are continuously removed from the system by interaction with the stellar radiation and the possible stellar wind. If the planetesimals are made not only of rock but contain also ice, their collisions and erosion can lead to the liberation of different gas molecules \citep{zuckerman2012,kral2017}. The lifetime of the released gas molecules in debris discs is limited due to rapid photodissociation in the absence of large amount of opaque dust grains. Considering only UV photons from the interstellar radiation field (ISRF), the photodissociation 
lifetime of unshielded CO molecules is only $\sim$120\,yr \citep{visser2009,heays2017}. 
With such a short lifetime, only the gas content of discs with lower CO masses could be explained within the framework of secondary gas disc models.

However, as recent studies 
demonstrated \citep{kral2019,cataldi2020,marino2020}, assuming 
sufficiently high 
(but still realistic) gas production rates from the icy bodies, neutral atomic 
carbon gas -- mainly produced via photodissociation of CO and CO$_2$ -- can become optically thick for UV photons that otherwise would dissociate CO molecules. 
This extra shielding by the carbon gas prolongs the photodissociation lifetime of 
CO significantly \citep{rollins2012}, allowing larger CO masses to accumulate in debris discs.   
At sufficiently high densities, self-shielding of CO gas becomes important too, further increasing the lifetime of the CO molecules in these optically thin discs.
A detailed modelling of these shielding processes shows that even the secondary gas scenario may work for explaining the formation of the CO-rich debris discs \citep[e.g.][]{kral2019,moor2019,marino2020}.

Debris discs emerge after the dispersal of Class\,II protoplanetary discs 
which are made of gas-rich primordial material.
Considering that all known CO-rich debris discs are young and probably represent the very early phase of debris disc evolution, \citet{kospal2013} raised an alternative scenario  \citep[see also][]{pericaud2017,nakatani2021} that CO-rich debris discs actually have a hybrid nature, where secondary debris dust and long-lived residual primordial gas from the preceding Class\,II disc phase coexist.
In this model, the necessary shielding of CO is related to the presence of 
leftover H$_2$ molecules. 
Should such hybrid discs exist, it would imply that during the transition from protoplanetary to debris disc the evolution of the gas and dust components 
can be decoupled from each other.
The origin of gas in the CO-rich debris discs is still under debate,
with recent observational pieces of evidence suggesting that in some discs, the gas could indeed be of second generation  \citep[e.g.,][]{hughes2017,kral2017,kral2019}.   

One possibility to decide which of the above scenarios works would be to investigate the chemistry in the CO-rich discs.
Whatever the origin of the gas in the CO-rich debris disc is, CO may not be the only abundant gas component. In younger, less-evolved Class~II protoplanetary discs a number of major C-, O-, N-, S-bearing polyatomic molecules have also been detected \citep[e.g.][]{Dutrey_ea14,Oberg_Bergin21,Pegues_ea21a}. In our Solar System, the primitive bodies such as comets have partly retained the primordial volatile matter left from its formation epoch. The cometary ices are mainly made of water, carbon monoxide, and carbon dioxide. In addition to these constituents, about two dozen other 
molecules (e.g. CH$_4$, C$_2$H$_2$, H$_2$CO, HCN) with 
lower abundances have been detected in 
cometary atmospheres \citep{bockelee-morvan2017}. 
The gas mixture of a secondary disc is expected to contain the same species, complemented
with their photodissociation products. The relative proportions of the individual constituents depend on the ice composition of the local exocomets, the mechanism that leads to the gas release, the UV radiation environment, 
and the shielding efficiency.

Contrary to this, the gas mixture in a hybrid disc is thought to be dominated by the long-lived primordial H$_2$ molecules. In addition, there could 
be different primordial residual gas species as well as secondary 
gas components released from the icy comets situated in the disc.
The presence of the large amount of H$_2$ in such a disc can 
lead to a significantly different chemical environment than in 
an H$_2$-poor secondary gas disc. 

Detecting other molecules than CO and thus better understanding the composition of the gas in debris discs can help us to elucidate the origin of the gas and, if the gas proves to be of secondary origin, to constrain the ice abundance of exocomets. Motivated by these opportunities, there already have been several attempts to survey the molecular content of gas-bearing debris discs. 
\citet{matra2018} used the ALMA interferometer and the Submillimeter Array (SMA) to search for several molecules (including CN, HCN, and {\hcop}) toward $\beta$\,Pic, the nearest known gaseous debris disc.  \citet{kral2020} targeted HD\,121191 and HD\,129590 with the ALMA interferometer to look for CN molecules. Finally, recently \citet{klusmeyer2021}
presented a deep molecular survey of the debris disc around 49\,Cet. They have targeted five molecules (HCN, CN, {\hcop}, SiO, and CH$_3$OH) with ALMA to characterise the molecular chemistry. These millimeter surveys resulted in non-detections for all species.The favourable edge-on or nearly edge-on orientation of discs around $\beta$~Pic and 49\,Cet allows to probe the gas material by measuring absorption lines in their optical/ultraviolet spectra 
\citep{hobbs1985,roberge2014}. Though, the application of this technique for $\beta$~Pic resulted in the detection of many different atomic species, even in these studies, only CO was the only detected molecule \citep[e.g.][]{vidalmadjar1994,roberge2000,roberge2006}. Moreover, the 
explored material is more related to regions located much closer to the star.
 
The four objects mentioned above can be classified as low or medium CO mass systems in the known sample of gas-bearing debris discs. $\beta$\,Pic and HD\,129590 with their CO masses of $M_{\rm CO}\lesssim$10$^{-4}$\,M$_\oplus$ probably have pure secondary origin \citep{cataldi2018,matra2018,kral2020}.
HD\,121191 and 49\,Cet are more CO-rich, raising the possibility that their gas material may be primordial \citep{hughes2008,moor2017,moor2019}. Nevertheless, based on their estimated CO masses, $\sim$0.0025\,M$_\oplus$ for HD\,121191 \citep{moor2017,kral2020} and $\sim$0.01\,M$_\oplus$ for 49\,Cet \citep{moor2019}, even these systems are not the most CO-rich debris discs. 

Our present study focuses on the molecular inventory of four other CO-rich debris discs: HD\,21997, HD\,121617, HD\,131488, and HD\,131835 \citep{kospal2013,moor2017}. The estimated CO content of these discs is higher (0.02--0.1\,M$_\oplus$) than those discussed in the previous paragraph, making the hybrid scenario attractive in the case of these CO-rich discs.
Using deep ALMA observations, we search for five molecules (CN, HCN, {\hcop},  C$_2$H, and CS) in these debris discs, compare the gas composition with that seen in primordial discs around mature Herbig\,Ae stars, and perform chemical simulations to learn if the observed gas composition could be consistent with the hybrid disc hypothesis.

We discuss the target selection process in Section~\ref{sec:targetselection}. Then, we describe the observations and data reduction in Section~\ref{sect:obs}. The results of the observations and the measured line fluxes and their upper limits are presented in Section~\ref{sec:results}. In Section~\ref{sect:discussion} we are aiming to interpret the data and provide the details of our modelling strategy. The final conclusions are drawn in Section~\ref{sect:conclusion}.

\section{Targets and selected molecules} \label{sec:targetselection}

Observations of rare CO isotopologues (C$^{18}$O and $^{13}$CO) implied CO gas masses of $>$0.01\,M$_\oplus$ in six gas-bearing debris discs so far. Out of these systems, 49\,Cet was already surveyed for several molecules by ALMA \citep{klusmeyer2021}, while for HD\,32297, no CO isotopologue data were available at the time when our target list was compiled.
Therefore, the remaining four CO-rich debris discs, HD\,21997, HD\,121617, HD\,131488, and HD\,131835 were selected as targets of this study. All these discs are hosted by young A-type stars. HD\,121617, HD\,131488, and HD\,131835 are members of the $\sim$16\,Myr old \citep{pecaut2016} Upper Centaurus Lupus subgroup of the 
Scorpius-Centaurus (Sco-Cen) association, while HD\,21997 belongs to the somewhat older \citep[$\sim$42\,Myr,][]{bell2015} Columba association. 

 To place into context the line strengths measured in our survey we constructed a reference sample of primordial discs. Previous studies focusing on molecular composition of protoplanetary disk material mostly targeted discs around young low-mass stars \citep[e.g.,][]{oberg2010,oberg2011,guilloteau2016}. Although some systems with Herbig Ae hosts have also been included in these studies, these targets were selected from the most massive, most gas-rich objects. For comparison with our debris disc sample , therefore we observed the same molecules in four nearby ($<$160\,pc), older, less gas-rich primordial discs around Herbig Ae stars, HD100453, HD139614, HD142666, and HD145718.
These four objects also belong to different subgroups of the Sco-Cen association. Based on their ages ($8-10$\,Myr, Table~\ref{tab:targets}) 
these discs represent a late phase of the protoplanetary disc evolution and 
thus are deemed as direct predecessors of the younger CO-rich debris discs. 

Moreover, our sample includes HD\,141569, a disc around a $\sim$6\,Myr old Herbig\,Ae star. Although this system is likely younger than any other in our sample (including the four protoplanetary discs) and the star exhibits signs of accretion, the dust and CO gas content of the disc resembles more that of CO-rich debris discs \citep{moor2019,difolco2020}. As an explanation of the observed characteristics, a recent study of the system suggests that while the inner disc is rather protoplanetary alike, 
the outer disc region (our measurements mainly probe that) is more evolved and might act like a debris disc \citep{difolco2020}. This object, thus, likely represents an intermediate stage between the protoplanetary and debris disc phases \citep[e.g.,][]{hughes2018}. The availability of good quality $^{13}$CO and/or C$^{18}$O measurements in the literature or the ALMA archive was another relevant consideration in their selection.  The fundamental properties of the selected nine systems are summarised in Table~\ref{tab:targets}.

\input{targetsrevised}

In order to test the compositional similarities of CO-rich debris discs and protoplanetary discs, we selected a list of molecules 
for our study. We target five molecules, HCN, CN, {\hcop}, \cch{}, and CS, 
which are routinely 
observed in protoplanetary discs, and whose line emission is typically only slightly weaker than that 
of $^{13}$CO  \citep[e.g.,][]{oberg2010,guilloteau2013,guilloteau2016,bergner2019,miotello2019}. 
The same molecules could also be used to test the similarity of the composition to solar systems' comets.
In addition to the three most abundant components, H$_2$O, CO, and CO$_2$, spectroscopic observations of cometary comas revealed two dozen other molecules \citep{bockelee-morvan2017}. HCN is a commonly detected 
component whose relative abundance to water (the dominant cometary volatile) ranges between 0.08 and 0.25\% 
in the observed comets. CN molecules are also detectable in comets. However, contrary to HCN, they are not primary volatiles released from the nucleus but instead form in the coma e.g. via photodissociation of HCN \citep{mumma2011}. Due to its long photodissociation lifetime, which is barely behind that of CO molecules, 
CN is considered to be the most promising molecule for detection in secondary gas material after CO \citep{matra2018}. 
C$_2$H is also not a primary volatile, but can be produced via the photodissociation of C$_2$H$_2$, a molecule 
already detected in several comets \citep{bockelee-morvan2017}. 
The detection of {\hcop} is less common, mostly limited to one Solar system comet, the Hale-Bopp \citep{veal1997}, 
where it is suggested to be formed via chemical reactions in the coma \citep{lovell1997,milam2004}.

\section{Observations and Data Reduction}
\label{sect:obs}
\input{obslog}

\begin{figure*}
	\includegraphics[width=\textwidth]{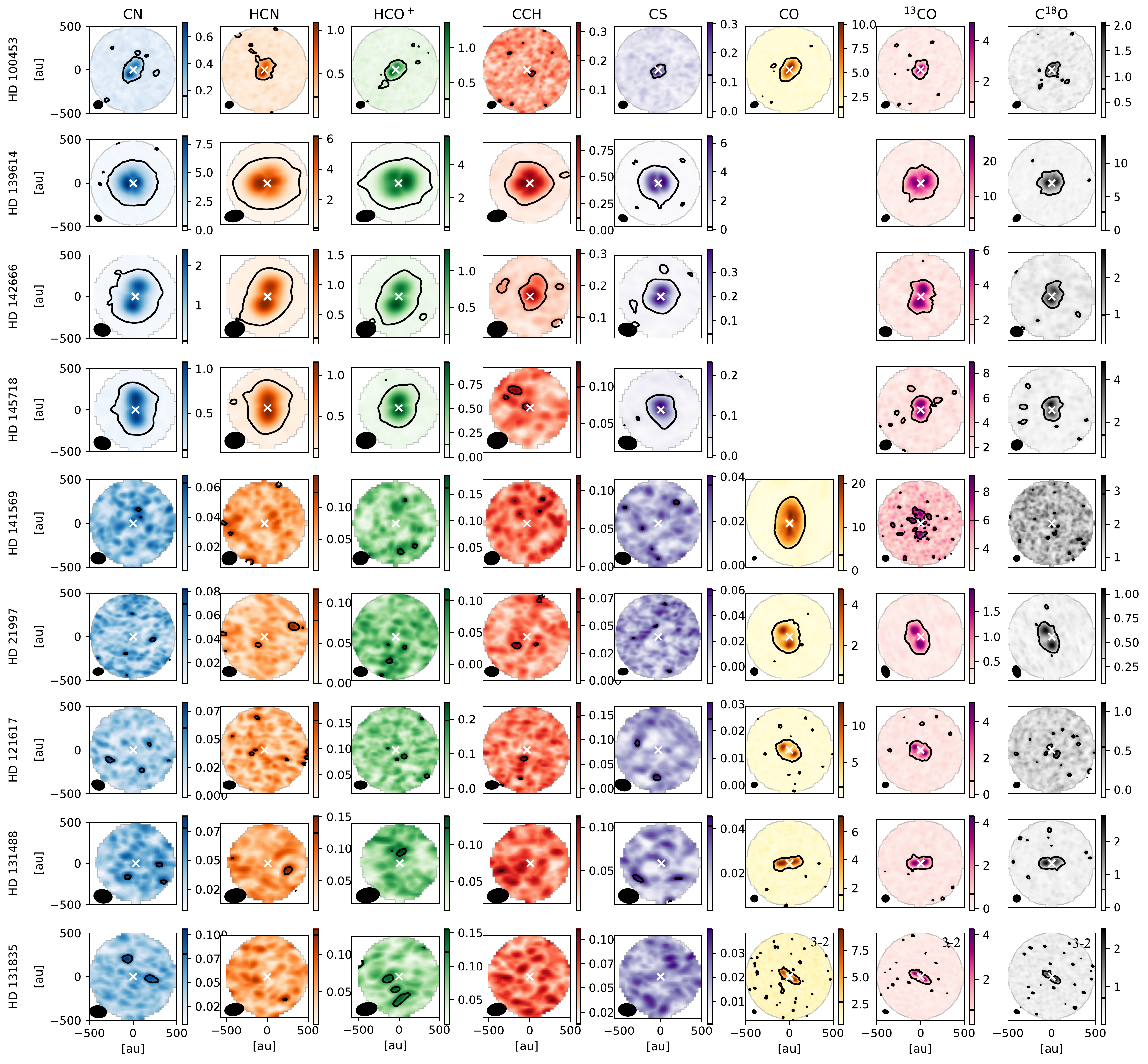}
    \caption{Peak brightness temperature for the observed discs. Contours show $3\sigma$ noise level. \hcop{}, HCN, CN, and \cch{} were observed within the project, and CO isotopologues represent archival ALMA data. Data for HD~141569 $^{12}$CO are taken from \citet{difolco2020}. HD~131835 is shown in CO isotopologues J=3-2 transitions, HD~141569 is in $^{12}$CO J=3-2, the others are in CO isotopologues J=2-1. The colour bar units are K.}
    \label{fig:peak}
\end{figure*}

Observations of our targets were performed with the ALMA 12m Array during 2018 and 2019 
in the framework of two projects (2017.1.01545.S and 2018.1.01429.S, PI: Th. Henning). 
The main parameters of these observations -- including their dates, the number of antennas, 
the baseline ranges of the array configurations, as well as the applied flux, bandpass, and phase calibrators -- are listed in Table~\ref{tab:obspar}. HD\,131488 and HD\,131835 as well as HD\,142666 and HD\,145718 are situated close to each other in the sky, allowing us to observe these pairs together in the same scheduling blocks.

Considering the typical low gas temperatures \citep{kospal2013,hughes2017} and the possible low density of potential collisional partners in CO-bearing debris discs, which can result in low excitation temperatures, we observed the lower transitions for the selected molecules. To observe the targeted lines, we defined one spectral setup in Band~5 and one in Band~6 for each object. Hyperfine transitions of the {\cch} N=2$-$1 line, as well as the J=2$-$1 lines of {\hcop} and HCN, were measured simultaneously in Band~5 using three spectral windows, each with a spectral resolution of 488.28\,kHz ($\sim$0.83{\kms}). This resulted in a total bandwidth of 937.5\,MHz (over 1920\,channels) in the spectral window, which targeted the {\cch} line. Centred at 174.96\,GHz, this window covered the hyperfine lines of {\cch}. The other two spectral windows at 177.27 and 178.2\,GHz have a total bandwidth of 468.75\,MHz (over 960\,channels) each. Two additional spectral windows with a bandwidth of 1.875\,GHz (128\,channels), centred at 187.5 and 189.4\,GHz were used to measure the Band\,5 continuum. In Band\,6 the correlator was set up with four spectral windows. Hyperfine transitions of CN N=2$-$1 were observed with a spectral resolution of 488.28\,kHz 
($\sim$0.65{\kms}) in a window centred at 226.65\,GHz. With the 1920 available channels, this provides a total bandwidth of 937.5\,MHz. 
Exploiting the capabilities of ALMA, in addition to the primarily targeted molecules, we also searched for additional molecular lines using the coarsest spectral resolution (976.56\,kHz, $\sim$1.2\,km~s$^{-1}$) in two basebands centred at 242.1 and 244.68\,GHz.
The fourth window, focused on measuring the dust continuum, has a central frequency of 229.00\,GHz and provides a bandwidth of 1.875\,GHz. The integration time ranged from 22 to 47 minutes per source, see Table~\ref{tab:obspar}. The frequencies of the transitions we used in our analysis are summarized in Table~\ref{tab:usedtransitions}.

\input{usedtransitions}

Calibration and flagging of the raw data sets were performed in the Common Astronomy Software 
Applications \citep[CASA, version 5.1.1,][]{mcmullin2007} using the standard ALMA reduction 
scripts. Prior to line imaging, we subtracted continuum emission from the obtained visibilities using the \textsc{uvcontsub} task of CASA. We fitted first-order polynomials to the line-free channels in the $uv$ space. Using the continuum-subtracted data, we applied the \textsc{tclean} task to construct spectral cubes of line observations. 
Natural weighting was adopted. The channel width was rebinned to
0.9{\kms} for {\cch}, {\hcop} and HCN lines and to 0.75{\kms} for CN observations, respectively. In the coarse-resolution Band\,6 measurements, a channel width of 1.25{\kms} was used.
While the remaining part of this paper focuses on the analysis of the molecular line results, 
for completeness in Appendix~\ref{appendix} we briefly summarise the processing and analysis 
of the continuum data.

\section{Results and analysis} \label{sec:results}

\begin{figure*}
	\includegraphics[width=\textwidth]{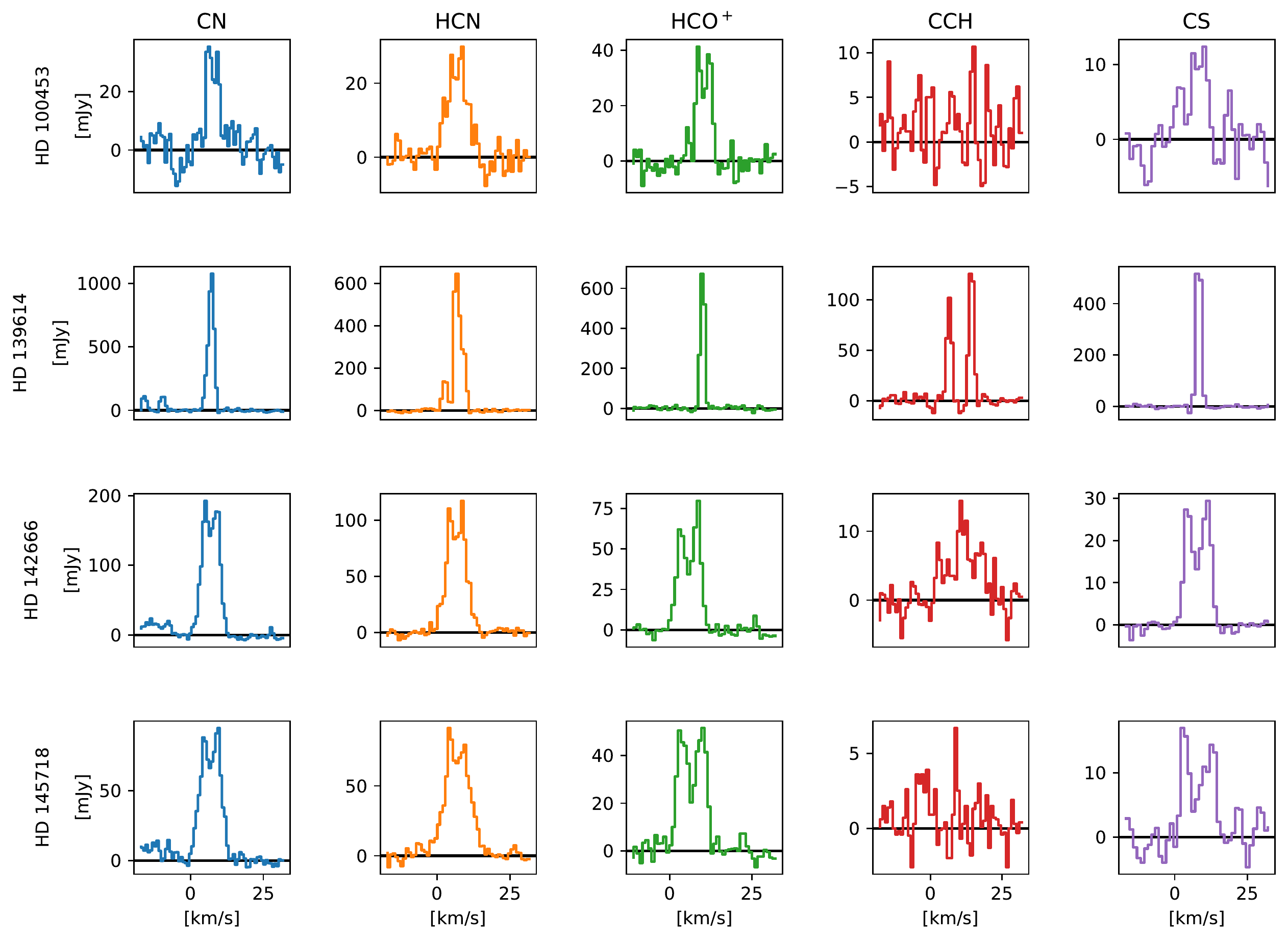}
    \caption{CN, HCN, \hcop{}, \cch{}, and CS emission line profiles of the observed Herbig disks.}
    \label{fig:spectra}
\end{figure*}

By inspecting the obtained data cubes, we found significant CN, HCN, 
and {\hcop} emission towards all four targeted HAe protoplanetary discs 
(HD\,100453, HD\,139614, HD\,142666, and HD\,145718), while 
the {\cch} line was detected only in HD\,139614 and HD\,142666 (Figs.~\ref{fig:peak},\,\ref{fig:spectra}). 
The J=$5-4$ transition 
of the CS molecule (Band\,6, 244.9355565\,GHz) was 
also detected in one of the additional coarse-resolution spectral windows  
in all four discs. 

In the detected cases, zeroth moment maps were constructed by integrating 
over the velocity ranges of the significant line emission.
The selected ranges are in good agreement with those inferred 
from the CO measurements of the same discs. The integrated line fluxes 
were derived by using elliptical apertures that were fitted to the 
observed velocity-integrated emission. To estimate flux uncertainties,
we randomly distributed 16 identical elliptical apertures in the background 
region and computed the standard deviation of the
fluxes obtained in them. 

In the case of CN and {\cch}, the 
spectral ranges cover several hyperfine transitions. For CN, the integrated 
maps, and thus the derived fluxes, are related to the $N=2-1$, 
$J=\frac{5}{2}-\frac{3}{2}$ transition, which consists of three blended hyperfine 
transitions ($F=\frac{7}{2}-\frac{5}{2}, \frac{5}{2}-\frac{3}{2}$, and
$\frac{3}{2}-\frac{1}{2}$), while for {\cch} the line fluxes are the sum 
of two hyperfine transitions ($F=3-2$ and $2-1$) of the $N=2-1$, 
$J=\frac{5}{2}-\frac{3}{2}$ line. 
No {\cch} emission was detected toward HD\,100453 and HD\,145718. 
In these cases, we computed upper limits for the integrated line flux. 
To this end, we determined the velocity range of the expected line
by using the HCN measurement of the given object as a template. 
We selected HCN because, on average, this line is the brightest among those detected for the given systems in our project. In determining the relevant velocity range, we also considered that the {\cch} is composed of two hyperfine transitions. Then we created zeroth moment maps. Finally, using the same elliptical aperture as in the corresponding HCN observations, we computed the 3$\sigma$ upper limits.

None of the four CO-rich debris discs, nor the disc around HD\,141569 
exhibited detectable line emission in any data cubes. 
For the calculation of the upper limits, we used the method described above, 
with the difference that we utilised the $^{12}$CO measurements of the 
given sources as a template. The CO measurements were taken from the ALMA archive 
and the template data cubes were smoothed using the \textsc{imsmooth} CASA task 
to obtain the same synthesised beam as that of the given line. 
We used the same channel width for these templates for the line measurements where upper limits
are to be determined. Thus, we used 0.9\,km~s$^{-1}$ for the {\cch}, {\hcop}, and HCN measurements 
and 0.75\,km~s$^{-1}$ for the CN observations. 
For the CN and {\cch} data, we considered the hyperfine structure, which is
resulted in broader velocity ranges to be used in the analysis.
In the case of HD\,21997, $\sim$50\% of the channels covering the {\hcop} line were flagged during the calibration process not to be used for scientific analysis. Therefore, this
upper limit estimate was obtained by slightly shifting the velocity range to a non-flagged 
part of the spectrum.
The measured integrated line fluxes are summarised in Table~\ref{tab:linefluxes}.

We present the resulting peak brightness maps in Fig.~\ref{fig:peak}. The first five rows are discs around Herbig Ae stars (HD~100453, HD~139614, HD~142666, HD~145718), followed by the more evolved disc HD~141569. The latter four rows show debris discs data (HD~21997, HD~121617, HD~131488, HD~131835). We also present moment zero maps in the Appendix (\ref{fig:mom0}).

Figure~\ref{fig:ratio} shows the measured CN, HCN, {\hcop}, {\cch}, and CS line flux ratios with respect to the $^{13}$CO (2$-$1) (left panel) and C$^{18}$O (2$-$1) (right panel) lines for our targets. Since for HD\,141569 
only an upper limit is available for the C$^{18}$O line flux, it is not shown in the corresponding panel. 
Apart from the case of C$_2$H measurements, the four protoplanetary discs are clearly separated from the debris discs and from HD\,141569 in these plots.
This is especially 
striking for the CN, HCN, and {\hcop} lines, even for HD\,100453  -- that exhibits the lowest line ratios 
-- the obtained ratios are at least 4$\times$ higher than the corresponding average upper limits of the debris disc sample. These results suggest that molecular abundances or physical conditions of the gas in the CO-rich debris discs is very different from those in the protoplanetary discs around Herbig\,Ae stars. We also find that HD\,141569 more resembles debris discs than our protoplanetary disc sample, similar to the previous results based on its dust and CO content (see Sect.~\ref{sec:targetselection}). 

\begin{figure*}
	\includegraphics[width=\linewidth]{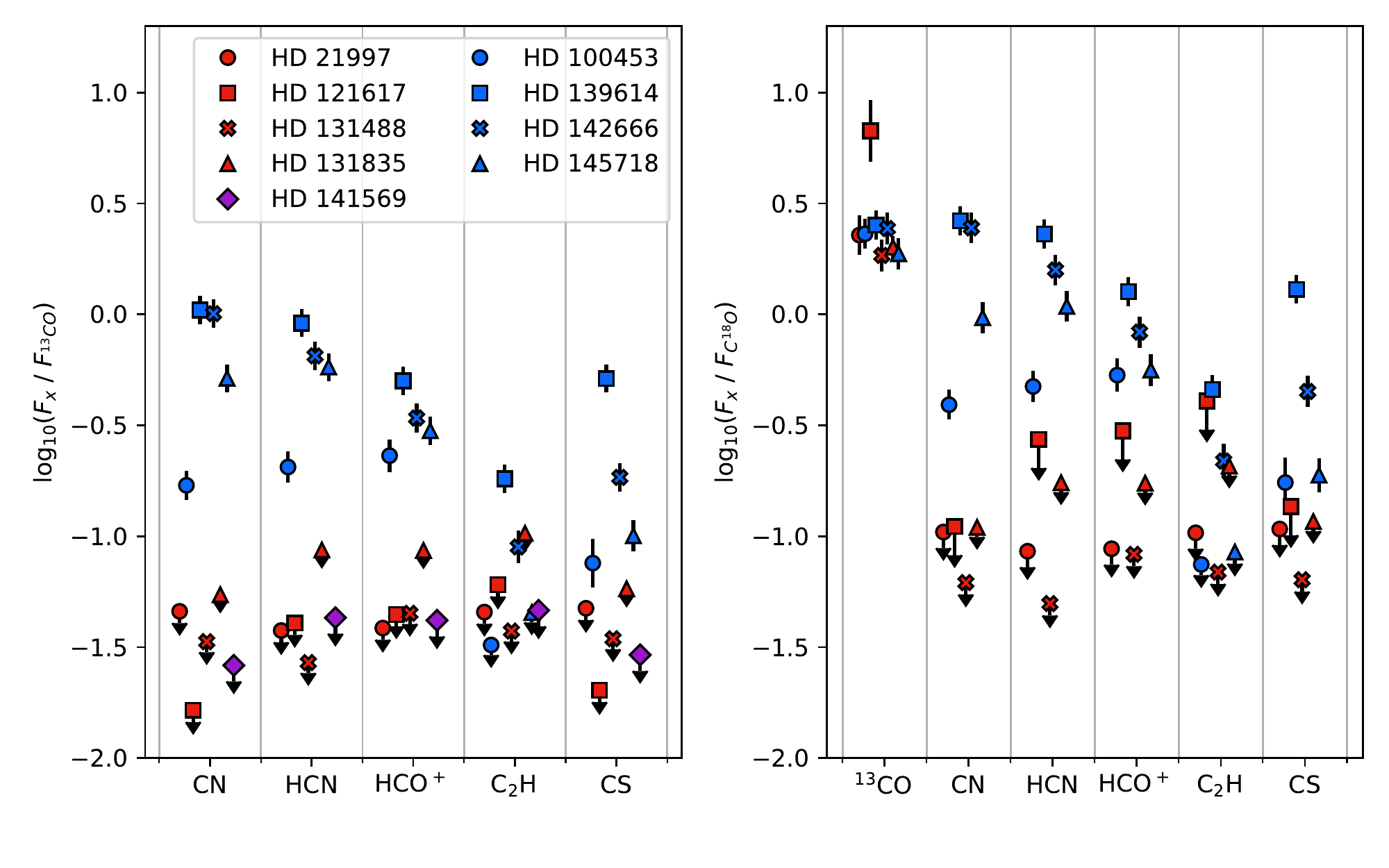}
	\caption{Line flux ratios of CN, HCN, {\hcop}, {\cch}, and CS molecules 
with respect to the $^{13}$CO 2$-$1 (left) and C$^{18}$O 2$-$1 (right) lines.  
The CN, HCN, {\hcop}, {\cch}, and CS line fluxes are from Table~\ref{tab:linefluxes}. 
The $^{13}$CO and C$^{18}$O data of HD 121617, HD\,131488, and HD\,21997 are from 
\citet{moor2017} and \citet{kospal2013}.  For HD\,131835, only the 3$-$2 
rotational transition of $^{13}$CO and C$^{18}$O has been measured (Mo\'or et al., 
in prep.) from which we estimated the 2$-$1 line fluxes. 
In the case of optically thick $^{13}$CO, the 
obtained 3$-$2 line flux was scaled by $(\nu_{3-2} / \nu_{2-1})^2 = 2.25$, 
while for the optically thin C$^{18}$O emission we 
assumed that local thermodynamical equilibrium holds
and that the gas temperature is 20\,K, resulting in a 3$-$2 
to 2$-$1 line flux ratio of 2.3.
As for the younger targets, CO flux data of HD\,100453 and HD\,141569 
are taken from \citet{vanderplas2019} and \citet{difolco2020}, respectively. 
For the remaining three objects, HD\,139614, HD\,142666, and 
HD\,145718, we inferred the necessary flux data from the analysis of
archival ALMA line observations (all belong to the project 2015.1.01600.S, PI: O. Panic).}
    \label{fig:ratio}
\end{figure*}

\section{Discussion}
\label{sect:discussion}
\subsection{Photodissociation of molecules}
Does the striking difference between the observed line ratios in the protoplanetary and the debris disc samples provide a clear hint that the gas in these debris discs is of secondary origin? Before examining this in more detail (Sect.~\ref{sect:model}), it is worth summarising how sensitive the studied molecules are to the UV radiation and which shielding mechanisms could protect them from photodissociation in an H$_2$-rich hybrid versus an H$_2$-poor secondary  environment. 

The photodissociation rate of a given molecule 
can be estimated as $k = k_0 \theta$, where $k_0$ is the unattenuated photodissociation 
rate of the molecule, while $\theta$ is the shielding function \citep{heays2017}. The latter function considers all other species that can shield the specific molecule against UV photons. An important difference between protoplanetary and debris discs is that optically thin dust cannot effectively attenuate the UV radiation in the debris discs.

\begin{figure}
	\includegraphics[width=\linewidth]{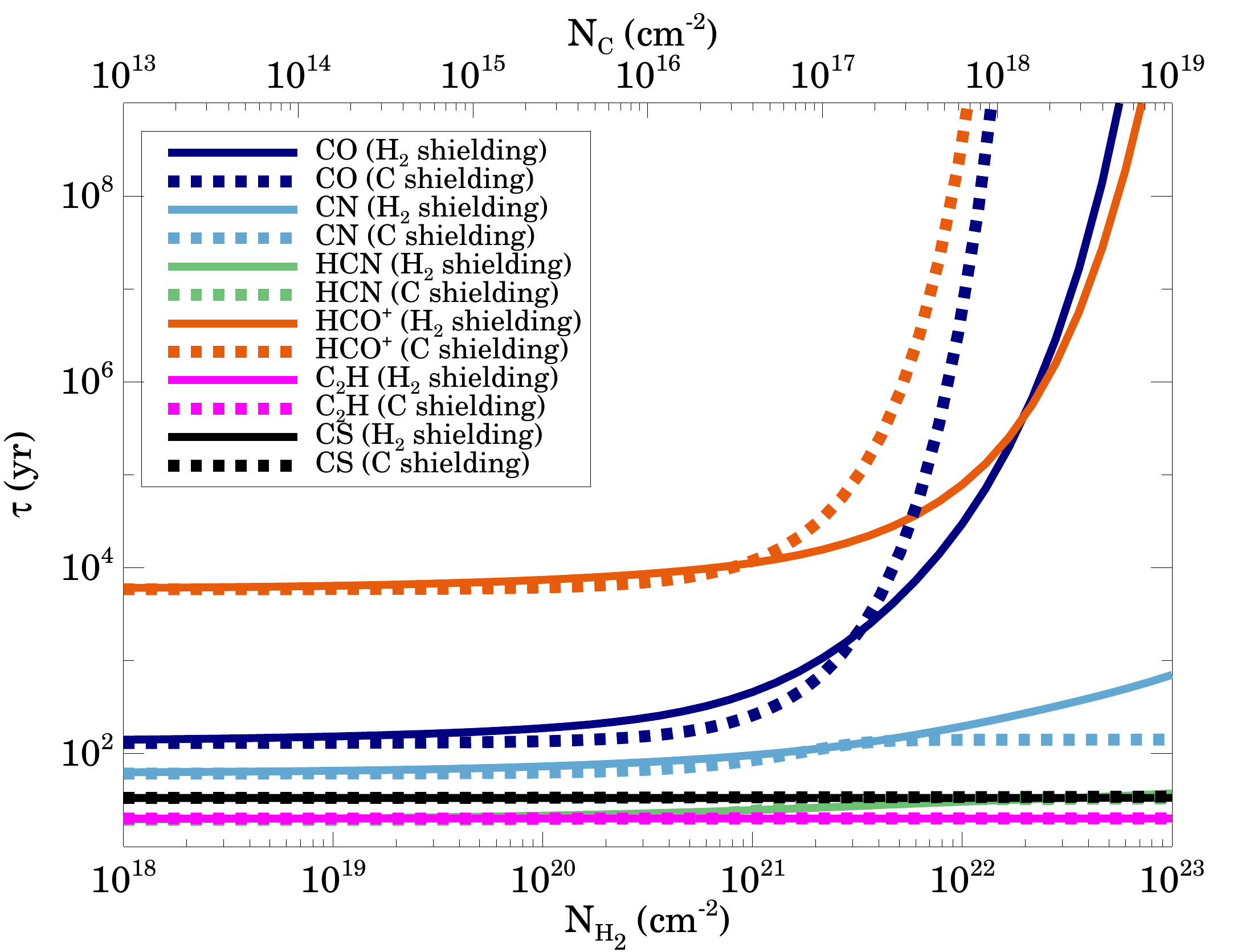}
	\caption{Photodissociation lifetimes of the CN, HCN, {\hcop}, C$_2$H, CS, and CO molecules 
in the interstellar radiation field, considering possible shielding by hydrogen 
molecules or carbon atoms.}
    \label{fig:pdlifetimes}
\end{figure}

In addition to the possible self-shielding of different molecules, in a hybrid disc dominantly H$_2$ molecules, while in secondary discs mainly C atoms can contribute substantially to the shielding \citep{kospal2013,kral2019}. 
Figure~\ref{fig:pdlifetimes} shows the 
photodissociation lifetimes ($\tau = 1/k$) of the targeted molecules and 
CO for the various H$_2$ or neutral atomic carbon column densities. We took into account only the interstellar UV radiation field in our calculations. 
The unattenuated photodissociation rates, as well as the corresponding shielding functions, 
were taken from \citet{heays2017}. 

As Fig.~\ref{fig:pdlifetimes} clearly demonstrates,
CO and {\hcop} molecules have quite long unattenuated photodissociation lifetimes, and could be efficiently shielded by both H$_2$ molecules and 
C atoms. In contrast, CN, HCN, C$_2$H, and CS  can absorb UV photons at wavelengths longward of 110\,nm, outside the wavelength interval where the H$_2$ molecules and C atoms can provide effective shielding. Without shielding at these long wavelengths, the latter four molecules have only short photodissociation lifetimes both in hybrid and shielded secondary discs, while in protoplanetary environments they could still be protected from UV by the optically thick dust. 
The observed low line ratios of these four species to CO 
in the debris disc sample, therefore, do not necessarily preclude the hybrid scenario (Fig.~\ref{fig:ratio}). In the case of more shielded {\hcop}, the longer photodissociation timescale and higher concentrations would be expected even in the optically thin debris discs. Using a physico-chemical and radiative transfer modelling, we examine below in more detail how the observed line ratios could constrain the nature of the CO-rich debris discs

\begin{table}
\caption{Basic properties of CO gas component in the debris disc subsample. 
The listed CO mass estimates are based on analysis of optically thin C$^{18}$O line observations.
References in the last column: 
(1): \citet{kospal2013}; 
(2): \citet{kral2019}; 
(3): \citet{moor2017}; 
(4): Pawellek et al. (in prep.). 
\label{tab:co_rings}
} 
\centering
\begin{tabular*}{\linewidth}{@{\extracolsep{\fill}}lcccc} 
\hline\hline
Object    & $R_{\textrm{in}}$, gas & $R_{\textrm{out}}$, gas & $M_{\textrm{CO}}$ & Ref.  \\
          & (au)                   & (au)       & (M$_\oplus$)      &       \\ 
\hline
HD 21997  & $<25$                  & 133             &   0.06   & 1   \\
HD 121617 & 50                     & 100             &   0.02   & 3     \\
HD 131488 & 35                     & 140             &   0.10   & 3,4    \\
HD 131835 & 50                     & 130             &   0.04   & 2,3   \\
\hline
\end{tabular*}
\end{table}

\input{linefluxtable}


\begin{figure*}
	\includegraphics[width=\textwidth]{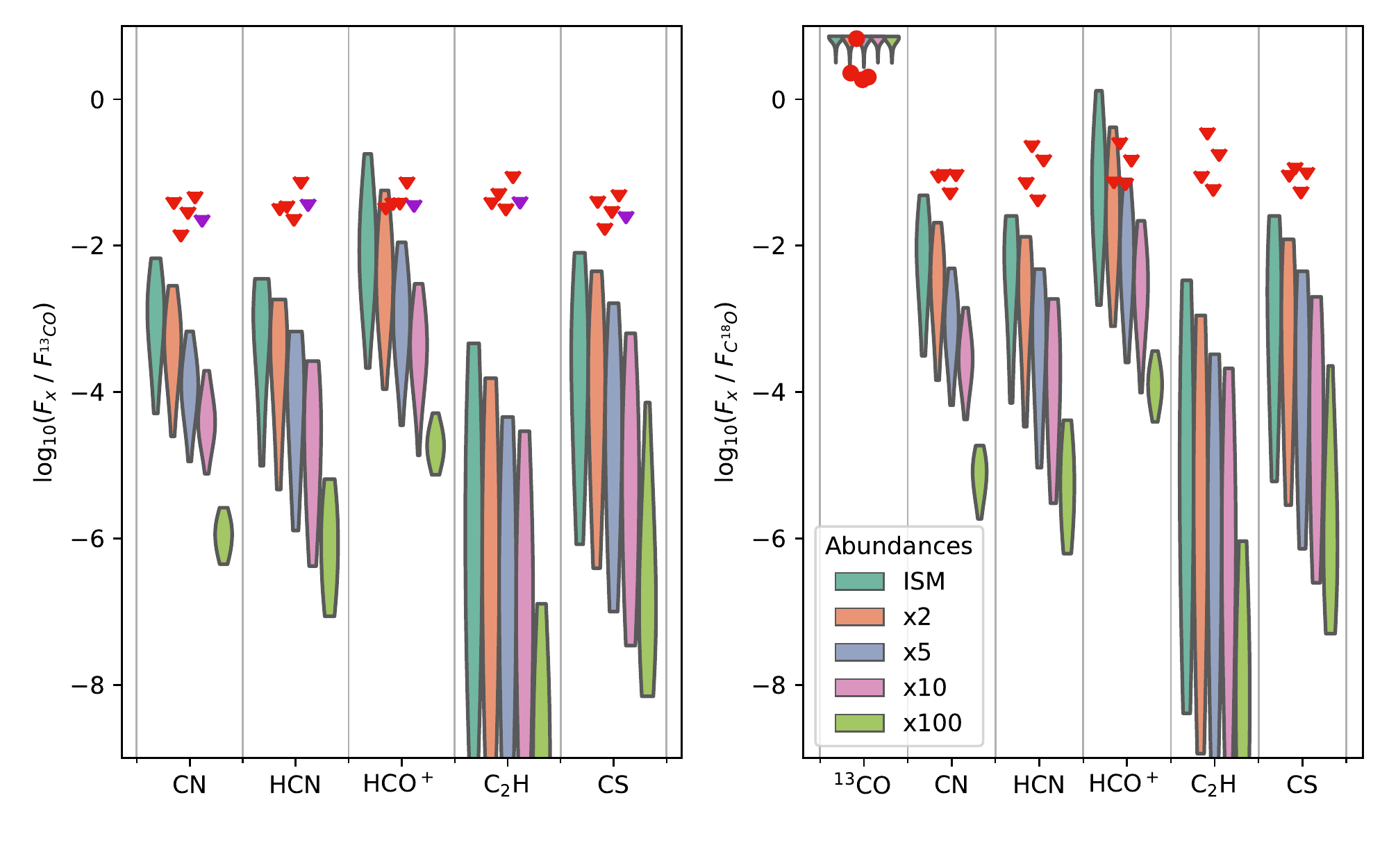}
    \caption{Scattered dots: observed line ratios, red –– for debris discs, purple –– for HD~141569. Circles are detections, and triangles are the upper limits. Violin bars: distributions of the modelled debris disc line ratios using primordial ISM-like and secondary hydrogen-depleted elemental ratios. Non-detection of CN, HCN, \cch, and CS in debris discs is fully reproduced. Significantly higher \hcop{} line fluxes are computed with the debris disc models using the H$_2$-rich, ISM-like initial abundances as compared with the H$_2$-poor case.}
    \label{fig:violin}
\end{figure*}

\subsection{Chemical and line flux modelling}
\label{sect:model}

\begin{table}
\caption{Initial Chemical Abundances. \label{tab:initial_abundances}}
\begin{tabular*}{\linewidth}{@{\extracolsep{\fill}}l|cc}
\hline\hline
            \noalign{\smallskip}
        Species & Abundance (ISM-like) & Abundance (H$_2$ and He are 100$\times$ depleted)\\
    & $n$(X) / $n$(H)  & $n$(X) / $n$(H) \\ 
            \noalign{\smallskip}
\hline
            \noalign{\smallskip}
H$_2$    & $5.00\ \times 10^{-1}$     &  $5.00\ \times 10^{-1}$    \\  
   He    & $9.75\ \times 10^{-2}$     &  $9.75\ \times 10^{-2}$    \\  
   C     & $7.86\ \times 10^{-5}$     &  $7.86\ \times 10^{-3}$    \\  
   N     & $2.47\ \times 10^{-5}$     &  $2.47\ \times 10^{-3}$    \\  
   O     & $1.80\ \times 10^{-4}$     &  $1.80\ \times 10^{-2}$    \\  
   S     & $9.14\ \times 10^{-8}$     &  $9.14\ \times 10^{-6}$    \\  
   Si    & $9.74\ \times 10^{-8}$     &  $9.74\ \times 10^{-6}$    \\  
   Mg    & $1.09\ \times 10^{-8}$     &  $1.09\ \times 10^{-6}$    \\  
   Fe    & $2.74\ \times 10^{-9}$     &  $2.74\ \times 10^{-7}$    \\  
   Na    & $2.25\ \times 10^{-9}$     &  $2.25\ \times 10^{-7}$    \\  
   Cl    & $1.00\ \times 10^{-9}$     &  $1.00\ \times 10^{-7}$    \\  
   P     & $2.16\ \times 10^{-10}$    &  $2.16\ \times 10^{-8}$    \\  
   \noalign{\smallskip}
\hline
\end{tabular*}
\end{table}

To reproduce detections of the CO isotopologues and non-detections of \hcop{}, CN, HCN, \cch, and CS in debris discs, we have utilised the ALCHEMIC chemical model \citep{Semenov_ea_2010} and the RADEX line radiative transfer code \citep{radex}, and adopted a temperature-density grid. Given that the observed CO emission appears as a relatively narrow ring-like structure in optically thin debris discs, we assume a rather tight variation of temperature and UV radiation strength within the gas emitting area. 
These physical conditions have been used to calculate time-dependent chemical evolution using the single-point ALCHEMIC chemical kinetics code and an up-to-date gas-grain chemical network (based on the KIDA'17 database with recent updates; \citet{KIDA, kida15}). The model includes gas-phase reactions, X-ray and UV photodissociation and photoionisation, cosmic ray-induced processes, as well as molecular freeze-out onto dust grain surfaces, surface reactions, and thermal and non-thermal desorption of ices. This chemical model is identical to the model used in \citet{2020A&A...644A...4S}, with the only exception of non-inclusion of deuterium, and hence it is only briefly summarised here. 

We have modelled chemical evolution over a temperature range between 10 and 300~K and densities between $10^3$ and $10^7$~cm$^{-3}$ (or between $\approx 4  \times  10^{-21}$ and $4 \times  10^{-17}$~g\,cm$^{-3}$). The unattenuated UV field from the star was set to be equal to the vertical ISRF \citep{1984ApJ...285...89D}. The absorption of UV photons was considered in the radial direction only, in a range between $\tau = 0$ and $\tau = 1$, which corresponds to the observed optical depth in debris discs, which are relatively dust-poor. 
Since primordial (sub-)micron-sized dust has already evolved and became severely depleted, we decided to set the corresponding dust-to-gas mass ratio to a value of $10^{-10}$ for $0.1$ $\mu$m-sized grains, effectively turning off dust-surface reactions. The ionisation rate was set to $1.3 \times 10^{-17}$ s$^{-1}$, corresponding to a galactic cosmic ray ionisation rate. High-energetic particle flux and X-ray radiation from the evolved main-sequence star were neglected, as these are much milder compared to the pre-main-sequence phase. Non-inclusion of X-ray and high energy particle radiation from the star does not affect the conclusions, as the higher intensity of ionising radiation in the model increases \hcop{}, thus increasing the area in the parameter space forbidden by the observational data.

We have considered five sets of initial abundances. The first so-called ``low metals'' set of mainly atomic abundances from \citet{Lee_ea98} describes the elemental content of the ISM that is available for gas or icy chemistry. It is often used to model protoplanetary discs, assuming a full chemical ``reset'' scenario  \citep{Eistrup_ea16,Drozdovskaya_ea16,2020A&A...644A...4S}. In the other four sets, we multiplied the relative abundances of all species other than hydrogen and helium by a factor of 2, 5, 10, and 100, respectively.
These four cases represent a scenario when a debris disc gas is partly replenished by secondary, collisionally-generated H$_2$-poor gas.  
Using the above chemical model, we have calculated the abundance evolution over 1~Myr for constant physical conditions. This timespan was long enough to reach a chemical equilibrium within 1-100~kyr for the observed, chemically simple molecules (depending on density). 
The final molecular concentrations have been used for the line radiative transfer simulations. 

\begin{figure}
\includegraphics[width=\columnwidth]{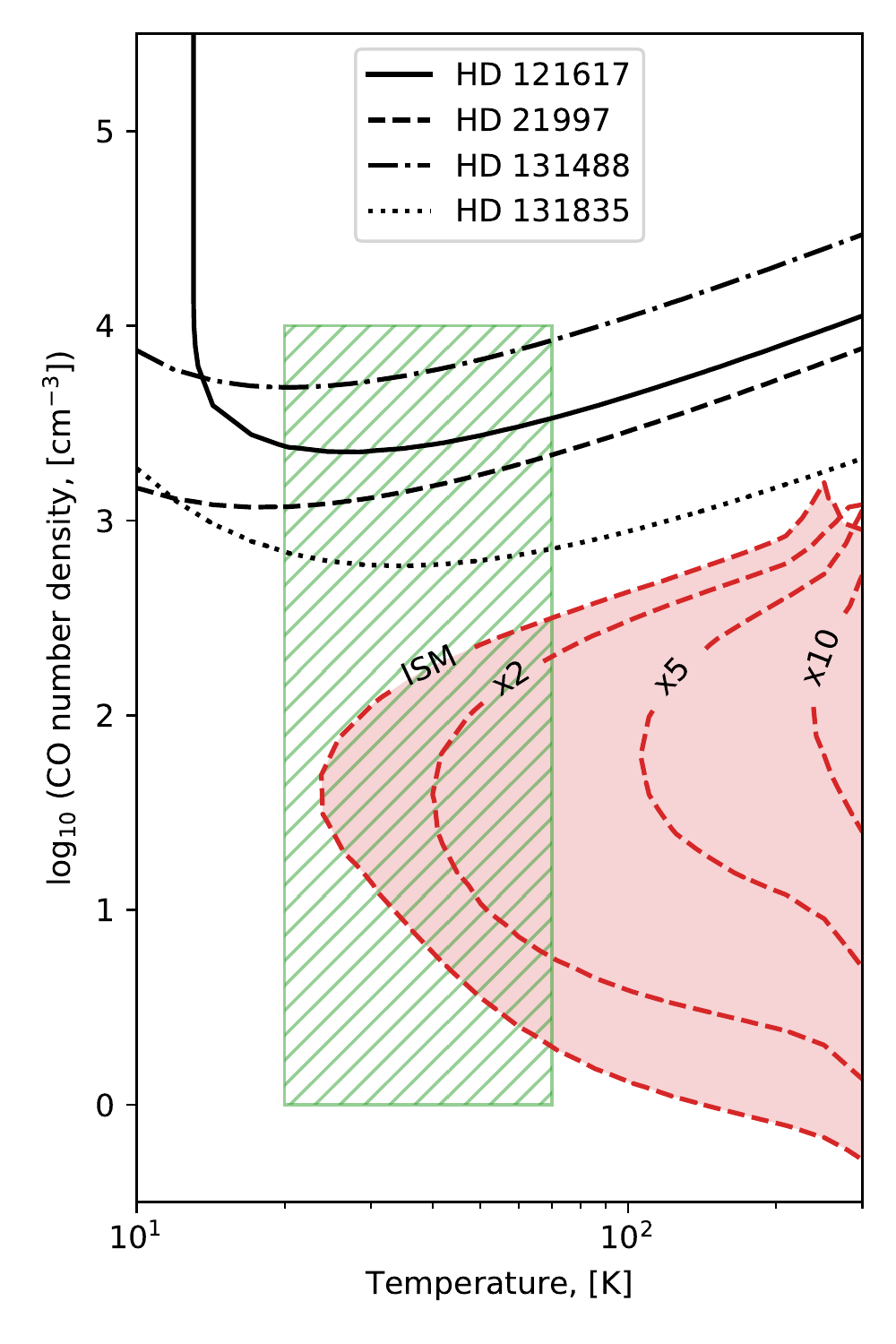}
    \caption{Modelled parameter space. Black lines: midplane densities at which the observed CO isotopologue brightness temperatures become fitted in the four debris discs. Green rectangle: a range of temperatures and densities representative of the debris discs. 
    Red filled contours: parameter combinations that lead to bright, detectable \hcop{} line emission (compared to the $^{13}$CO intensity), which contradicts the \hcop{} non-detections in the debris discs. Each red dashed contour corresponds to the different initial abundance sets. For the ISM-like or moderately H$_2$-depleted abundances, the low-density regions of the debris disc shall contribute to the \hcop{} emission, which could be detected by deeper ALMA observations. If H$_2$ is more severely depleted, by a factor of 5 or larger, \hcop{} molecules cannot form in sufficient amount under the debris disc conditions.}
    \label{fig:diagram}
\end{figure}

We have calculated the line brightness temperatures with the non-LTE RADEX code. The collisional cross-section and other spectroscopic data from the Leiden Atomic and Molecular Database (LAMDA) have been utilised \citep{lamda}. Since $^{13}$C and $^{18}$O are not included in the chemical network, we have used fixed isotopic ratios C$^{12}$/C$^{13} = 77$ and O$^{16}$/O$^{18} = 560$ to scale down the resulting molecular abundances \citep{wilson_rood_1994_isotope_ratio}. We have calculated the line brightness temperatures assuming the length of the emitting gas column of 15~au and a line full width at half maximum (FWHM) of 1~km\,s$^{-1}$. 
The collision partner densities have been taken from the chemical simulations. We have calculated the CN brightness temperature using hyperfine components from \citep{2015MNRAS.446.1750F}.

The results of the debris disc line simulations are presented in Fig.~\ref{fig:violin}. Using violin bars, we show the smoothed distribution of the simulated line ratios for the modelled disc cells.  We also overplot this figure with the observed line ratios and their upper limits, similar to Fig.~\ref{fig:ratio}. As it can be clearly seen from  Fig.~\ref{fig:violin}, all line intensities other than \hcop{} are expected to be below the detection limits, regardless of the initial abundances and position in the temperature-density parameter space. In the H$_2$-poor cases, the modelled CN, HCN, \hcop{}, \cch, and CS line intensities relative to the CO emission are predicted to decrease monotonically with the hydrogen depletion factor for the same range of the CO densities. 

We present the constraints on the parameter space for a single-point debris disc model in Fig.~\ref{fig:diagram}. We select a temperature range between 20 and 70~K and the CO volume density between $1$ and $10^4$~cm$^{-3}$. 
Our upper boundary to the debris disc gas density is calculated by dividing the measured $^{13}$CO and C$^{18}$O masses by the fitted area and the scale heights of the corresponding CO emitting rings estimated from the modelling (Table~\ref{tab:co_rings}). 
Disc midplane densities are shown in Fig.~\ref{fig:diagram} as black lines. With red contours, we show the regions in the parameter space, for which the predicted \hcop{} J=2-1/$^{13}$CO J=2-1 flux ratio is higher than the detection limit. 

The selected parameter space partially yields higher \hcop{} intensities than the upper limit set by our observations. That leaves a possibility that deeper observations could allow detecting \hcop{} in debris discs if their gas retains ISM-like composition and is H$_2$-rich. In such discs, 
the \hcop{} J=2-1 line would become detectable if the gas is warmer than 20~K and the CO density is about $3-300$ cm$^{-3}$.
If a debris disc possesses an extended tenuous atmosphere, the \hcop{} emission from this region  also contributes to the total \hcop{} line flux and hence increase the resulting \hcop{}/CO line ratio.
 With the currently available data, we estimate that the contribution of the debris disc atmosphere to the total \hcop{} line flux is negligible even in the H$_2$-rich ISM case. Follow-up higher-resolution CO or more sensitive \hcop{} observations combined with feasible thermo-chemical modelling are required to provide tighter constraints on the CO/H$_2$ ratios in these discs.

The alternative scenario, where the disc gas is already partially or fully depleted of H$_2$ shows lower \hcop{} abundances and hence the \hcop{} fluxes. Our model with the lowest H$_2$ depletion factor of 2 still produces enough \hcop{} to overcome the current detection limit at higher temperatures ($>40$ K) and volume densities about 80~cm$^{-3}$. However, starting from a higher H$_2$ depletion factor of 5, all the predicted \hcop{} J=2-1/$^{13}$CO J=2-1 flux ratios fall below the current detection limit.

Lower H$_2$ concentrations (with respect to CO) in the H$_2$-poor cases lead to the less efficient formation of the main driver of the gas-phase chemistry, H$_3^+$. This, in turn, makes abundances of all observed molecules apart from CO also lower by factors of at least several or higher. According to our KIDA-based model, H$_3^+$ is produced via H$_2$ + H$_2^+$ reaction driven by the cosmic ray ionization of H$_2$. Next, H$_3^+$ can react with CO and form \hcop{} and atomic H. H$_3^+$ can also react with atomic C and produce CH$^+$ and H$_2$. Another starting route for hydrocarbon chemistry is a slow radiative association reaction between C$^+$ and H$_2$ that leads to CH$_2^+$. Both CH$^+$ and CH$_2^+$ can react with H$_2$, forming bigger CH$_n^+$ ions ($n = 3-5$). These light hydrocarbon ions can be neutralised by dissociative recombination with electrons and then undergo carbon insertion reactions with C$^+$, forming larger C$_2$H$_n^+$ species ($n \geq 0$). These C$_2$-hydrocarbon ions can react with e$^-$ and form neutral C$_2$-hydrocarbons, such as C$_2$H molecule targeted in our observations.

Furthermore, hydrocarbon chemistry is also a key for the formation of nitriles via ion-molecule or neutral-neutral reactions involving atomic N. A key ion-molecule reaction is CH$_3^+$ + N $\rightarrow$ HCNH$^+$ + H, followed by dissociative recombination of HCNH$^+$ into CN or HCN or HNC (with equal probability). Slower barrierless neutral-neutral reactions between N and either CH or CH$_2$ produce H and either CN or HCN/HNC. Thus, the lack of H$_2$ in the scenario when debris disc has secondary origin could explain the overall deficit of molecular species other than CO.

\subsection{Effect of hydrogen depletion on disc scale height}

An alternative method to estimate hydrogen depletion of a debris disc using ALMA is based on direct measurements of their vertical temperature and density structure \citep{hughes2017, kral2019}. In the case of hydrostatic equilibrium, the vertical scale height is defined by the molar mass and temperature. In the simplest case of an isothermal and geometrically thin disc atmosphere, scale height $H$ is defined as:

\begin{equation}
    H = \sqrt{\frac{RT}{\mu} \frac{r^3}{GM}},
\end{equation}
where $R$ is the ideal gas constant, $T$ is the gas temperature, $\mu$ is the gas molar mass, $r$ is the radial distance to the star, $G$ is the gravitational constant, and $M$ is the stellar mass. 
The measurements of the scale height were performed below, for example, for the edge-on debris disc around 49~Ceti \citet{hughes2017} estimated that the disc scale height is smaller by 25\% compared to the value predicted from the temperature measurements and assuming the ISM-like, H$_2$-rich gas.

In the case of hydrogen depletion, the gas molar mass becomes higher, as it is not dominated by the molecular hydrogen and helium any more. The molar mass of a gas mixture could be calculated as:
\begin{equation}
    \mu = \frac{\sum_i{X_i}\mu_i}{\sum_i{X_i}},
\end{equation}
where $X_i$ is an abundance of the $i$-th species, and $\mu_i$ is its molar mass.

\begin{figure}
	\includegraphics[width=\columnwidth]{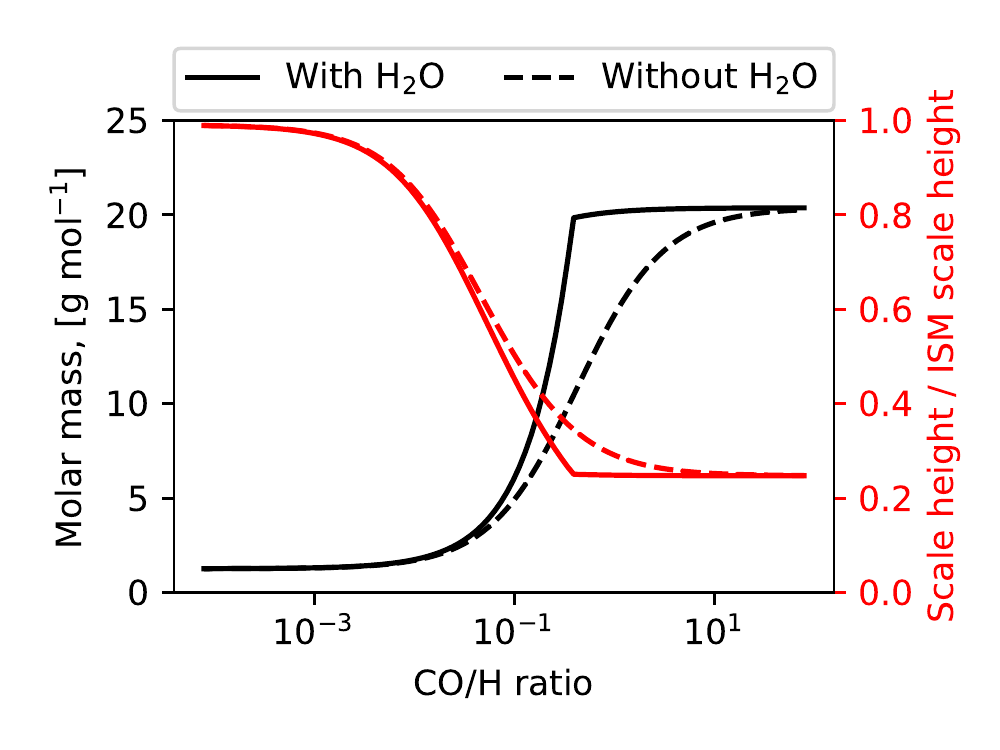}
    \caption{Gas molar mass (black) and disc scale height (red) as a function of relative CO abundance with respect to hydrogen. 
    Solid line: gas consisting of CO, H$_2$O and H$_2$. Dashed line: gas consisting of CO, OH, and H$_2$.}
    \label{fig:mmw}
\end{figure}

To illustrate this effect, we use the elemental abundances from Table~\ref{tab:initial_abundances} and scale abundances of all the species other than molecular hydrogen and helium by uniform factors of $1$ (ISM-like) to $10^6$ (hydrogen is almost fully depleted). The resulting elemental abundances are then reshuffled into various major molecules, assuming simplistic chemistry. First, all available carbon is assumed to be locked in CO molecules. Then, the remaining oxygen and available hydrogen are assumed to form water molecules. After that, the remaining hydrogen (if any) is assumed to be bound in H$_2$ molecules. As water can be rapidly photodissociated in the optically thin debris disc, we consider two extreme cases -- with and without water formation. The resulting gas molar masses and the corresponding scale heights for an isothermal, 30~K disc are shown in Fig.~\ref{fig:mmw}. The effect of hydrogen depletion on the disc scale height becomes noticeable as soon as the CO/H ratio is above $\sim 10^{-2}$. For example, the scale height measured in the 49~Ceti disc by \citep{hughes2017} would correspond to the CO/H ratio of about $2\times 10^{-2}$ in this case. 


If there is a vertical temperature gradient, the hydrostatic equilibrium and the corresponding ideal gas equation are as follows:
\begin{equation}
\begin{cases}
    \frac{dP(z)}{dz} = -\rho(z) g_z = -\rho(z) \frac{GMz}{(r^2 + z^2)^{3/2}}\\
    P(z) =\frac{\rho(z) RT(z)}{\mu}
\end{cases} 
\end{equation}

By differentiating the ideal gas equation and assuming that the gas molar mass $\mu$ is constant, one derives:
\begin{equation}
    \frac{dP(z)}{dz} = \frac{R}{\mu}\left(T(z)\frac{d\rho(z)}{dz} + \rho(z)\frac{dT(z)}{dz}\right)
\end{equation}

Combining the above equations, we can get the equation to measure the local gas molar mass directly:



\begin{equation}
    \mu = - \frac{R(r^2+z^2)^{3/2}}{GMz}\left(T(z)\frac{d\ln{\rho(z)}}{dz} + \frac{dT(z)}{dz}\right)
\label{eq:mmw}
\end{equation}

If vertical density $\rho(z)$ and temperature $T(z)$ profiles could be reliably measured from the observations, the gas molar mass could also be derived from this Eq.~\ref{eq:mmw}. This would require further high-resolution ALMA observations of the multi-J CO lines in edge-on-oriented debris discs, as has been tried for 49~Ceti.

\section{Conclusions}
\label{sect:conclusion}
We present ALMA Band~5 and 6 observations of four debris discs (HD~21997, HD~121617, HD~131488, HD~131835), and five old Herbig Ae discs (HD~139614, HD~141569, HD~142666, HD~145718, HD~100453). We run the standard CASA pipeline with natural weighting to produce continuum images and emission lines data cubes. We have measured fluxes in elliptical apertures and estimated the noise by comparison with other apertures at other locations of the data cube. 
All discs were detected in the dust continuum and CO isotopologues, while other targeted lines (CN, HCN, \hcop{}, \cch{}, and CS) were only detected in the Herbig Ae discs. The upper limits of the flux ratios to CO isotopologue lines in debris discs were found to be significantly lower than the ratio in Herbig Ae discs, {except for \cch{}, which we did not detect in all the Herbig Ae discs, and thus do not have strict constrain.

We performed detailed chemical modelling using a parameter grid to find physical conditions and elemental abundance combinations that could reproduce this behaviour. 
We found that in the optically thin debris discs, where molecules other than CO and H$_2$ are rapidly photodissociated, \hcop{} is the only relatively abundant and potentially detectable molecule. This requires the emitting gas to be warmer than 20~K and the CO density to be low, about $3-300$ cm$^{-3}$. 
In the scenario when the debris disc is of secondary origin and is hydrogen-poor,  the predicted molecular abundances and hence line fluxes are significantly lower. In this case, the calculated line fluxes scale with the degree of the hydrogen depletion. 

We also propose that higher-resolution studies of the vertical distribution of the debris disc gas could be used to constrain its elemental composition. The lack of hydrogen and helium in the scenario when the gas is collisionally produced should increase the gas molar mass above the ISM-like value of 2.3~g\,mol$^{-1}$, resulting in the noticeable decline in the local scale height. 

The spatial resolution and reached sensitivity of our ALMA observations did not allow us to verify the exact origin of the gas in the studied four debris discs. 
Thus, future detection(s) of \hcop{} or the CO data obtained at higher resolution are needed to provide a clearer view on the nature of the debris disc gas.

\section*{Acknowledgements}
This paper makes use of the following ALMA data: ADS/JAO.ALMA\#2017.1.01545.S 
and 2018.1.01429.S. ALMA is a partnership of ESO (representing its member states), 
NSF (USA) and NINS (Japan), together with NRC (Canada) and NSC and ASIAA (Taiwan) 
and KASI (Republic of Korea), in cooperation with the Republic of Chile. The Joint 
ALMA Observatory is operated by ESO, AUI/NRAO and NAOJ. This work has made use of 
data from the European Space Agency (ESA) mission{\it Gaia} (\url{https://www.cosmos.esa.int/gaia}), 
processed by the {\it Gaia} Data Processing and Analysis Consortium (DPAC,
\url{https://www.cosmos.esa.int/web/gaia/dpac/consortium}). Funding for the DPAC
has been provided by national institutions, in particular the institutions
participating in the {\it Gaia} Multilateral Agreement. This research has made use 
of the VizieR catalogue access tool, CDS, Strasbourg, France (DOI : 10.26093/cds/vizier). 
The original description of the VizieR service was published in A\&AS 143, 23. A.M. and P. \'A. acknowledge support from the grant KH-130526 of the Hungarian NKFIH.
D.S. acknowledges financial support by the Deutsche Forschungsgemeinschaft through SPP 1833:
``Building a Habitable Earth'' (SE 1962/6-1). A.M.H. is supported by a Cottrell Scholar Award from the Research Corporation for Science Advancement. This work was supported by ''Programme National de Physique Stellaire'’ and "'Programme Physique et Chimie du Milieu Interstellaire"(PNPS and PCMI from INSU/CNRS).
Authors acknowledge the impact of the open-source Python community, and, particularly, the developers and maintainers of Astropy, NumPy, Matplotlib, seaborn, pandas.

\section*{Data Availability}

The data underlying this article are available in the ALMA Science Archive and can be accessed with (2017.1.01545.S and 2018.1.01429.S, PI: Th. Henning).



\bibliographystyle{mnras}
\bibliography{bib}




\appendix
\section{Results of the continuum observations} \label{appendix}

Our observations enabled us to characterise the continuum emission of the nine discs at 1.27\,mm (in Band~6) and 1.65\,mm (in Band~5). 
Although the data analysis is performed in the visibility space, 
for completeness, continuum images of the discs are also 
presented in Fig.~\ref{fig:cont}.
We used the 
\textsc{uvmultifit} package \citep[ver.~3.0,][]{marti-vidal2014} to fit 
the obtained visibility data in the \textit{uv} plane by a geometrical model. Only channels free from line emission were considered. The data weights in the measurement set were updated with the \textsc{statwt} task before the fitting process.
We adopted an elliptical Gaussian model, where the axis ratio (computed from the inclinations) 
and the position angle ($PA$) were fixed based on literature data (see 
Table~\ref{tab:contdiskpars}) leaving 
four free parameters: the positional offsets with respect to the 
phase center, the total flux density of the fitted component ($F_\nu$), and 
the full width half maximum (FWHM) of the major axis. In the case of HD\,121617 
 the axis ratio and the $PA$ were also fitted because these parameters are 
 less constrained in the literature than for the other 
 targets. After the fitting, we used the CASA \textsc{tclean} task to 
 image the residuals obtained as the difference of the measurement and 
 best fit model. We found that the images of HD\,141569 display an 
 extended residual emission implying the presence of an additional 
 broader disc component. This is in good agreement with the findings 
 of previous high-resolution millimeter continuum observations 
 of the source \citep{white2018,miley2018,difolco2020}. 
With this in mind, for this target, we used a two-component model. 
For the Band~5 observation, where the central component could not 
be resolved, a combination of a point source and an elliptical Gaussian 
was adopted, while the shorter wavelength data was fitted by combining 
two elliptical Gaussian components. 

Comparing the centre of the fitted models with the 
Gaia\,EDR3 positions of the stars (corrected for proper motion) we found 
no significant offsets. 
The obtained flux densities and disc sizes (FWHMs of the major axes) 
with their uncertainties are presented in Table~\ref{tab:contdiskpars}. 
The quoted uncertainties of the flux densities in parentheses are 
quadratic sums of the measurements errors and absolute calibration errors.
The latter component is conservatively assumed to be 10\%\footnote{According to 
the ALMA Technical handbook (\url{https://arc.iram.fr/documents/cycle6/ALMA_Cycle6_Technical_Handbook.pdf}) 
the accuracy of the flux calibration is 
$\leq$10\% in these bands.}.

\begin{figure}
	\includegraphics[width=0.95\linewidth]{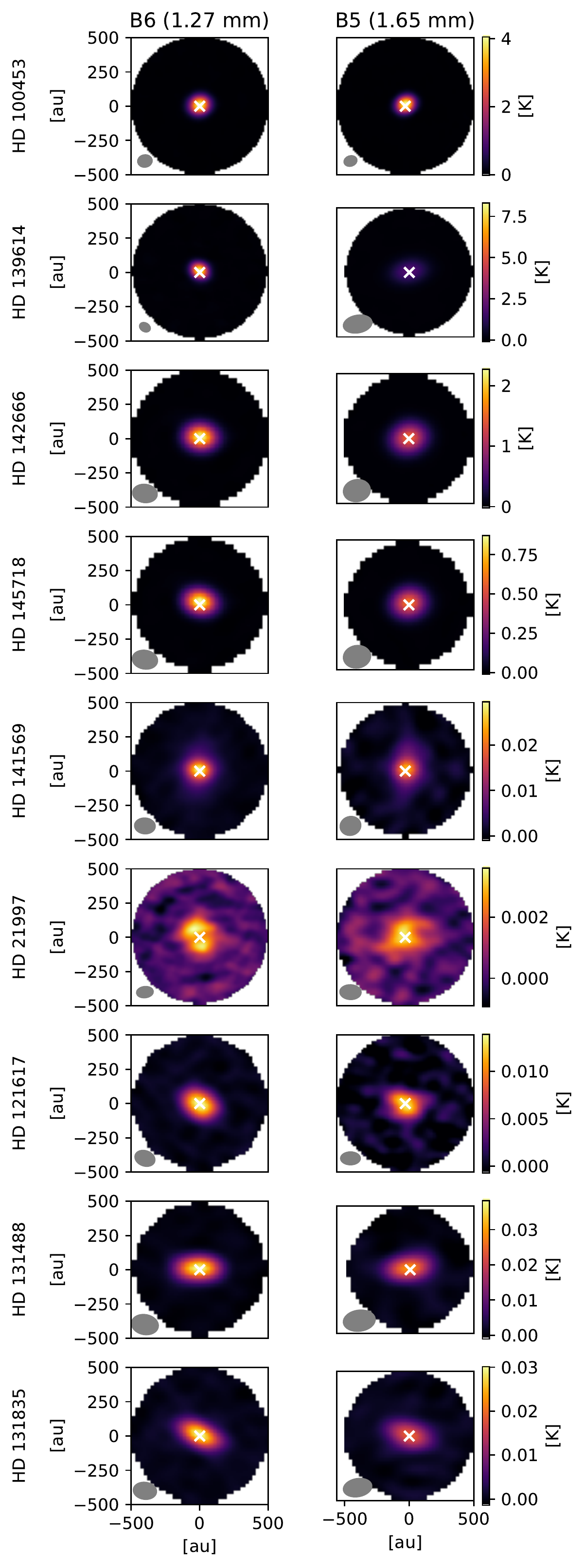}
    \caption{ALMA Band 5 and Band 6 continuum emission for the observed discs. To produce these images we used the CASA task tclean function with a Briggs robust parameter of 0.5 and considered data from all line-free channels. The colourbar units are K.}
    \label{fig:cont}
\end{figure}
\input{contdiskpars}



\section{Moment zero maps}

We also provide the moment zero maps (Fig.~\ref{fig:mom0}) of the same data cubes as on peak brightness maps (Fig.~\ref{fig:peak}).

\begin{figure*}
	\includegraphics[width=\textwidth]{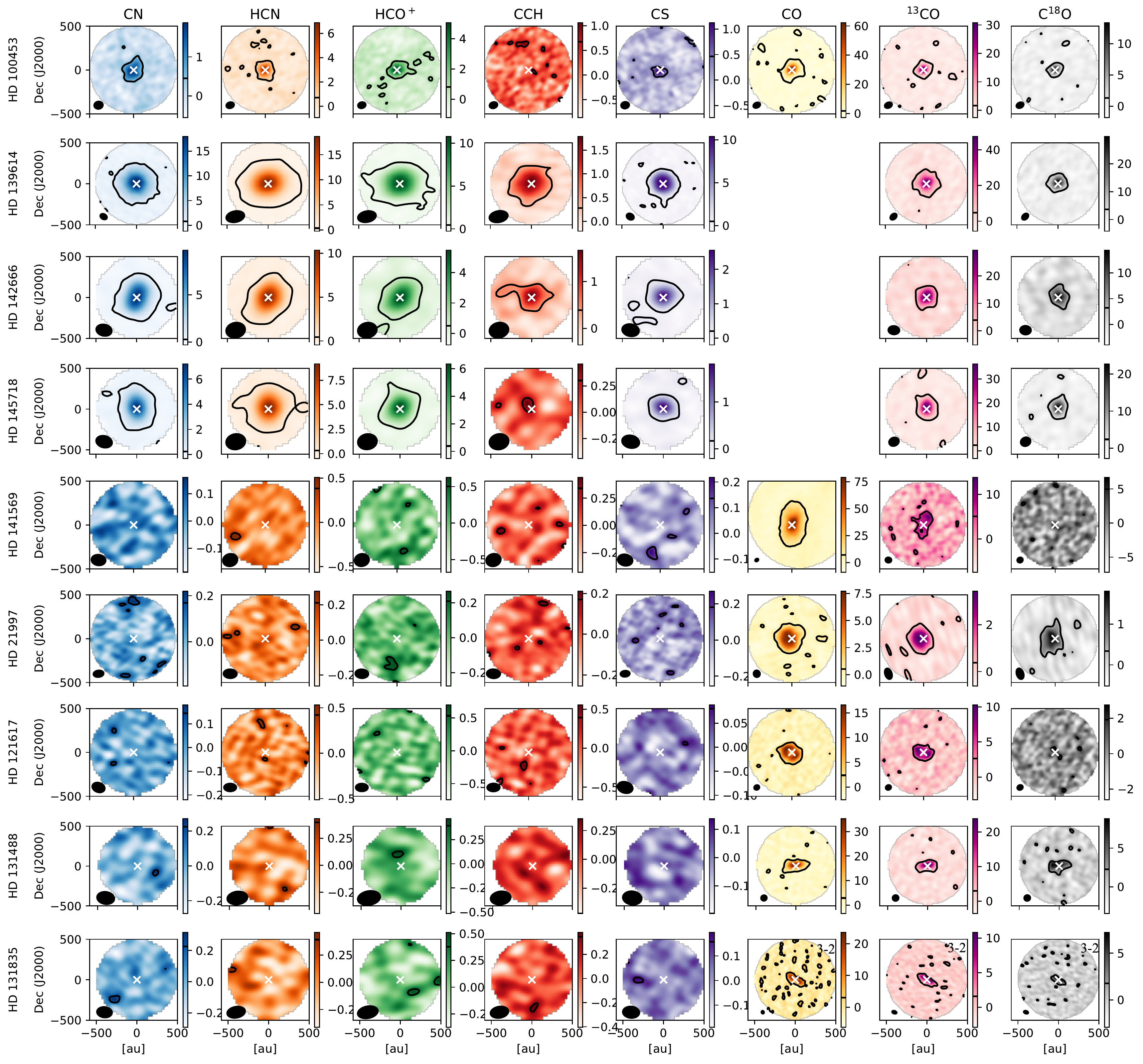}
    \caption{Moment zero maps for the observed discs. HD~131835 is shown in CO isotopologues J=3-2 transitions, the other are in CO isotopologues J=2-1. The velocity range was selected based on CO emission data cubes. The colourbar units are K km / s. Data for HD~141569 $^{12}$CO are taken from \citet{difolco2020}.}
    \label{fig:mom0}
\end{figure*}

\bsp	
\label{lastpage}
\end{document}

%% file: targetsrevised.tex
\begin{table*}                                                                  
\begin{center}                                                                                                                                       
\caption{Overview of the sample. Distance estimates are taken from \citet{bailerjones2021}. Group membership information is taken from 
  the literature \citep{bell2015,dezeeuw1999,hoogerwerf2000}, abbreviations in this column are as follows 
  -- COL: Columba moving group; LCC: Lower Centaurus Crux association; UCL: Upper Centaurus Lupus association; 
  US: Upper Scorpius association. 
  For debris systems we quote the age of the group to which the given object belongs, while for Herbig\,Ae objects 
  individual age estimates are listed from \citet{murphy2021} and \citet{wichittanakom2020}. Fractional luminosities 
  ($L_\mathrm{disk}/L_\mathrm{*}$) of the debris discs are taken from \citet{moor2017}, while similar data of 
  the proplanetary sample are from \citet{dent2005} and \citet{meeus2012}.      
  The listed $^{13}$CO ($L_{\rm ^{13}CO}$) and C$^{18}$O luminosities ($L_{\rm C^{18}O}$) refer to the J=2$-$1 rotational 
  transition, except for HD\,131835 (marked by asterisk) where the J=3$-$2 data are quoted. 
  For HD\,139614, HD\,142666, and
HD\,145718 the CO luminosity data are derived from analysis of
archival ALMA line observations obtained in the framework of ALMA project 2015.1.01600.S (PI: O. Panic). For the rest 
of the sample, including both the remaining Herbig stars and the debris discs, the luminosity values are computed based on CO line fluxes, obtained by the ALMA or NOEMA arrays and  
taken from the literature \citep[][Mo\'or et al. in prep.]{difolco2020,kospal2013,moor2017,vanderplas2019}     
  {References used in the last column:} 
  (1): \citet{bailerjones2021}; 
  (2): \citet{bell2015}; 
  (3): \citet{dent2005};
  (4): \citet{dezeeuw1999}; 
  (5): \citet{difolco2020}; 
  (6): \citet{hoogerwerf2000}; 
  (7): \citet{kospal2013};
  (8): \citet{meeus2012}; 
  (9): \citet{moor2017}; 
  (10): Mo\'or et al. in prep.
  (11): \citet{murphy2021}; 
  (12): \citet{pecaut2016}; 
  (13): \citet{vanderplas2019}; 
  (14): \citet{wichittanakom2020}.
 \label{tab:targets} }
\begin{tabular}{lccccccccccc}                                                     
\hline\hline
Target name & Spectral Type & Distance              & Group & $T_{\rm eff}$ & $L_*$       & $M_*$       &  Age  & $L_\mathrm{disk}/L_\mathrm{*}$ & $L_{\rm ^{13}CO}$  & $L_{\rm C^{18}O}$   &  Ref. \\
            &               &  (pc)                 &       &   (K)         & (L$_\odot$) & (M$_\odot$) & (Myr) &    &  (10$^{17}$W)    &  (10$^{17}$W)      &           \\             
\hline
HD\,21997   &  A3V          &  69.6$\pm$0.1         & COL   & 8300 & 10.4 & 1.75 & 42   & 5.7$\times$10$^{-4}$  & 3.5  & 1.5    & 1,2,7,9 \\
HD\,121617  &  A1V          & 117.5$\pm$0.5         & UCL   & 9050 & 14.9 & 1.92 & 16   & 4.8$\times$10$^{-3}$  & 6.4  & 1.0    & 1,6,9,12 \\
HD\,131488  &  A1V          & 151.4$^{+0.6}_{-0.8}$ & UCL   & 9000 & 13.5 & 1.88 & 16   & 5.5$\times$10$^{-3}$  & 9.8  & 5.3    & 1,6,9,12 \\
HD\,131835  &  A2V          & 129.1$^{+0.5}_{-0.4}$ & UCL   & 8400 & 10.3 & 1.75 & 16   & 3.0$\times$10$^{-3}$  & 11.5$^{*}$ & 5.8$^{*}$ & 1,4,9,10,12 \\
HD\,141569  &  B9.5V/A0Ve   & 111.3$^{+0.3}_{-0.4}$ & -     & 9500 & 21.9 & 2.06 & 5.9  & 9.0$\times$10$^{-3}$  & 13.0 & $<$1.1 & 1,5,8,14 \\
HD\,100453  & A9Ve          & 103.7$\pm$0.2         & LCC   & 7250 & 7.1  & 1.48 & 11.0 & 6.2$\times$10$^{-1}$  & 8.5  & 3.7    & 1,4,8,13,14  \\
HD\,139614  & A7Ve          & 133.1$^{+0.5}_{-0.4}$ & UCL   & 7650 & 6.7  & 1.52 & 10.8 & 3.9$\times$10$^{-1}$  & 41.4 & 16.4   & 1,6,8,11 \\
HD\,142666  & A8Ve          & 145.5$\pm$0.4         & US    & 7250 & 12.0 & 1.64 & 7.8  & 3.3$\times$10$^{-1}$  & 23.9 & 9.8    & 1,6,8,14  \\
HD\,145718  & A5Ve          & 153.9$^{+0.5}_{-0.4}$ & US    & 7750 & 11.2 & 1.64 & 8.7  & 1.0$\times$10$^{-1}$  & 29.1 & 15.5   & 1,3,6,14 \\ 		   
\hline
\end{tabular}
\end{center}
\end{table*}

%% file: obslog.tex
\begin{table*}                                                                  
\begin{center}                                                                                                                                       
\caption{Observational parameters \label{tab:obspar} }
\begin{tabular}{lccccccccc}                                                     
\hline\hline
Target name & Project code & Obs. date & Antennas \# & Baselines (m) & Band & \multicolumn{3}{c}{Calibrators} \\
            &              &           &                    &               &      & Flux        &   Bandpass   &      Phase  & Int. time (m) \\    
\hline
HD\,21997  & 2018.1.01429.S & 2019-03-20&  44                & 15--314       &   5  & J0423-0120  & J0423-0120   &  J0329-2357 & 34 \\
HD\,21997  & 2018.1.01429.S & 2019-03-11&  44                & 15--314       &   6  & J0423-0120  & J0423-0120   &  J0329-2357  & 31 \\
HD\,100453 & 2018.1.01429.S & 2018-12-06&  45                & 15--784       &   5  & J1107-4449  & J1107-4449   &  J1132-5606 & 27 \\
HD\,100453 & 2018.1.01429.S & 2018-12-19&  45                & 15--500       &   6  & J1107-4449  & J1107-4449   &  J1132-5606 & 24 \\
HD\,121617 & 2017.1.01545.S & 2018-05-28&  43		     & 15--457       &   5  & J1427-4206  & J1427-4206   &  J1321-4342 & 47 \\
HD\,121617 & 2017.1.01545.S & 2018-08-18&  45		     & 15--314       &   6  & J1427-4206  & J1427-4206   &  J1424-4913 & 42 \\
HD\,131488 & 2017.1.01545.S & 2018-07-09&  44		     & 15--314       &   5  & J1517-2422  & J1517-2422   &  J1457-3539 & 35 \\
HD\,131835 & 2017.1.01545.S & 2018-07-09&  44		     & 15--314       &   5  & J1517-2422  & J1517-2422   &  J1457-3539 & 35 \\ 
HD\,131488 & 2017.1.01545.S & 2018-07-16&  43		     & 15--314       &   6  & J1427-4206  & J1427-4206   &  J1457-3539 & 32 \\
HD\,131835 & 2017.1.01545.S & 2018-07-16&  43		     & 15--314       &   6  & J1427-4206  & J1427-4206   &  J1457-3539 & 31 \\
HD\,139614 & 2017.1.01545.S & 2018-08-24&  45		     & 15--500       &   5  & J1617-5848  & J1617-5848   &  J1555-4150 & 25 \\
HD\,139614 & 2017.1.01545.S & 2018-04-01&  44		     & 15--704       &   6  & J1427-4206  & J1427-4206   &  J1604-4441 & 22 \\
HD\,141569 & 2017.1.01545.S & 2018-08-24&  43		     & 15--314       &   5  & J1550+0527  & J1550+0527   &  J1557-0001 & 25 \\
HD\,141569 & 2017.1.01545.S & 2018-04-01&  42		     & 15--284       &   6  & J1517-2422  & J1517-2422   &  J1557-0001 & 22 \\
HD\,142666 & 2017.1.01545.S & 2018-07-04&  44		     & 15--314       &   5  & J1517-2422  & J1517-2422   &  J1551-1755 & 24 \\
HD\,145718 & 2017.1.01545.S & 2018-07-04&  44		     & 15--314       &   5  & J1517-2422  & J1517-2422   &  J1551-1755 & 24 \\
HD\,142666 & 2017.1.01545.S & 2018-05-30&  43		     & 15--314       &   6  & J1517-2422  & J1517-2422   &  J1551-1755  & 22 \\
HD\,145718 & 2017.1.01545.S & 2018-05-30&  43		     & 15--314       &   6  & J1517-2422  & J1517-2422   &  J1551-1755  & 22 \\
\hline
\end{tabular}
\end{center}
\end{table*}

%% file: usedtransitions.tex
\begin{table}                                                                  
\begin{center}                                                                                                                                       
\caption{List of the transitions used in our analysis. Note that in the case of C$_2$H and CN our observations covered more transitions but those were not utilized in our work. 
Rest frequencies are taken from 
the Splatalogue database (\url{https://www.cv.nrao.edu/php/splat/}). \label{tab:usedtransitions}}

\begin{tabular}{lcc}                                                     
\hline\hline
Molecule & Transition & Rest frequency (GHz) \\
\hline
C$_2$H  & $N=2-1$,$J=\frac{5}{2}-\frac{3}{2}$,$F=3-2$ & 174.6632220 \\
        & $N=2-1$,$J=\frac{5}{2}-\frac{3}{2}$,$F=2-1$ & 174.6676850 \\
HCN     & $J=2-1$ & 177.2611115 \\
{\hcop} & $J=2-1$ & 178.3750563 \\
CN      & $N=2-1$,$J=\frac{5}{2}-\frac{3}{2}$,$F=\frac{5}{2}-\frac{3}{2}$ & 226.8741908 \\
        & $N=2-1$,$J=\frac{5}{2}-\frac{3}{2}$,$F=\frac{7}{2}-\frac{5}{2}$ & 226.8747813 \\
        & $N=2-1$,$J=\frac{5}{2}-\frac{3}{2}$,$F=\frac{3}{2}-\frac{1}{2}$ & 226.8758960 \\
CS      & $J=5-4$ & 244.9355565 \\
\hline
\end{tabular}
\end{center}
\end{table}

%% file: linefluxtable.tex
\begin{table*}
\begin{center}
\caption{Integrated line fluxes and 3$\sigma$ upper limits for our targets. Uncertainties in parentheses
are quadratic sums of the measurement errors and the overall calibration uncertainty (assumed
to be 10\%). \label{tab:linefluxes} }
\begin{tabular}{lccccc}
\hline\hline
Target name  & $S_{\rm CN}$ & $S_{\rm HCN}$ & $S_{\rm HCO^+}$ & $S_{\rm CCH}$ & $S_{\rm CS}$  \\
             & (mJy~km~s$^{-1}$) & (mJy~km~s$^{-1}$) & (mJy~km~s$^{-1}$) & (mJy~km~s$^{-1}$) & (mJy~km~s$^{-1}$) \\
\hline

  HD 100453 &            153$\pm$6 (16) &           185$\pm$12 (22) &           208$\pm$19 (28) &                   $<$29.1 &      69.5$\pm$16.3 (17.7) \\
  HD 139614 &         2790$\pm$47 (283) &         2433$\pm$32 (245) &         1338$\pm$23 (136) &           484$\pm$16 (51) &         1369$\pm$19 (138) \\
  HD 142666 &         1296$\pm$15 (130) &           836$\pm$15 (85) &           439$\pm$12 (46) &           115$\pm$10 (15) &            237$\pm$7 (25) \\
  HD 145718 &           721$\pm$19 (75) &           811$\pm$14 (82) &           419$\pm$16 (45) &                   $<$63.4 &           141$\pm$11 (18) \\
  HD 141569 &                   $<$31.4 &                   $<$51.5 &                   $<$50.1 &                   $<$55.6 &                   $<$35.0 \\
     HD 21997 &                   $<$37.6 &                   $<$30.8 &                   $<$31.6 &                   $<$37.3 &                   $<$38.8 \\
  HD 121617 &                    $<$8.7 &                   $<$21.5 &                   $<$23.5 &                   $<$32.0 &                   $<$10.7 \\
  HD 131488 &                   $<$16.4 &                   $<$13.2 &                   $<$22.0 &                   $<$18.3 &                   $<$16.9 \\
  HD 131835 &                   $<$12.7 &                   $<$20.2 &                   $<$20.1 &                   $<$24.0 &                   $<$13.5 \\
\hline
\end{tabular}
\end{center}
\end{table*}

%% file: contdiskpars.tex
\begin{table*}                                                                  
\begin{center}                                                                                                                                       
\caption{Continuum results. Flux densities and disc sizes (FWHMs of the major axes) are derived by fitting the obtained visibility data. Apart from the case of HD 121617, where these parameters are also estimated from our measurements, inclination and positional angle data are taken from the literature (for references see the last column). \label{tab:contdiskpars} }
\begin{tabular}{lcccccc}                                       
\hline\hline
Target name & Wavelength & F$_\nu$ & FWHM       & $i$  & $PA$ & Ref. \\
            &  (mm)  &  (mJy)  & (arcsec)  & (deg.) & (deg.) &  \\             
\hline
HD\,21997   & 1.27 & 1.88$\pm$0.14 (0.23) & 3.3$\pm$0.2 & 32.9 & 21.5 &  \citet{moor2013} \\
            & 1.65 & 1.21$\pm$0.17 (0.21) & 3.4$\pm$0.4 &  &  &   \\
HD\,121617  & 1.27 & 1.67$\pm$0.05 (0.17) & 1.34$\pm$0.06 & 43.9$\pm$3.0  & 53.5$\pm$4.0 &   \\
            & 1.65 & 0.92$\pm$0.05 (0.10) & 1.42$\pm$0.13 & 43.4$\pm$8.6 &  59.8$\pm$12.3 &   \\
HD\,131488  & 1.27 & 2.86$\pm$0.04 (0.29) & 1.17$\pm$0.05 & 82 & 96 & \citet{moor2017}  \\
            & 1.65 & 1.64$\pm$0.04 (0.17) & 1.10$\pm$0.11 &  &  &   \\
HD\,131835  & 1.27 & 2.59$\pm$0.04 (0.26) & 1.44$\pm$0.05 & 79 & 59 & \citet{kral2019}   \\
            & 1.65 & 1.40$\pm$0.05 (0.15) & 1.53$\pm$0.10 &  &  &   \\
HD\,141569  & 1.27 & 1.55$\pm$0.10 (0.18) & 0.39$\pm$0.12 & 53 & 356 & \citet{difolco2020}  \\
            &      & 2.47$\pm$0.17 (0.30) & 3.51$\pm$0.20 &  &  &   \\ 
            & 1.65 & 0.74$\pm$0.07 (0.10) & - &  &  &   \\
	    &      & 1.66$\pm$0.15 (0.22) & 3.25$\pm$0.31 &  &  &   \\ 
HD\,100453  & 1.27 & 218.1$\pm$1.3 (21.8)  & 0.64$\pm$0.01 & 29.5 & 151.0 & \citet{vanderplas2019}  \\
            & 1.65 & 110.9$\pm$0.8 (11.1)  & 0.65$\pm$0.01 &      &       &   \\
HD\,139614  & 1.27 & 191.5$\pm$0.6 (19.2) & 0.50$\pm$0.01 & 17.6 & 276.5 & \citet{muro-arena2020}  \\
            & 1.65 &  98.1$\pm$0.7 (9.8) & 0.49$\pm$0.01 &      &       &   \\
HD\,142666  & 1.27 & 120.3$\pm$0.2 (12.0) & 0.41$\pm$0.01 & 62.2 & 162.1 & \citet{huang2018}  \\
            & 1.65 & 67.9$\pm$0.1 (6.8) & 0.44$\pm$0.01 &  &  &   \\
HD\,145718  & 1.27 & 44.8$\pm$0.7 (4.5) & 0.30$\pm$0.01 & 70.4 & 1.0 & \citet{ansdell2020}  \\
            & 1.65 & 26.0$\pm$0.3 (2.6) & 0.28$\pm$0.03 &  &  &   \\	        	        	    	    
\hline
\end{tabular}
\end{center}
\end{table*}